# Network-Level Cooperative Protocols for Wireless Multicasting: Stable Throughput Analysis and Use of Network Coding

Anthony Fanous, *Student Member, IEEE,* and Anthony Ephremides, *Life Fellow, IEEE*

**Abstract**

In this paper, we investigate the impact of network coding at the relay node on the stable throughput rate in multicasting cooperative wireless networks. The proposed protocol adopts Network-level cooperation in contrast to the traditional physical layer cooperative protocols and in addition uses random linear network coding at the relay node. The traffic is assumed to be bursty and the relay node forwards its packets during the periods of source silence which allows better utilization for channel resources. Our results show that cooperation will lead to higher stable throughput rates than conventional retransmission policies and that the use of random linear network coding at the relay can further increase the stable throughput with increasing Network Coding field size or number of packets over which we encode.

## I. INTRODUCTION

Cooperative Diversity was shown to be effective in combating multipath fading over wireless channels as it enables single antenna users to benefit from spatial diversity achieved by a relay node or possibly another source node. Much work has been done to analyze cooperative diversity at physical layer and based on information theoretic considerations [1], [2]. It has also been shown that cooperative diversity can be implemented at the network layer by using the fact the traffic is bursty and hence allowing the relay to use the idle time slots of the source without according any explicit resources to the relay. In [3] a network level cooperative protocol has been shown to increase the stable throughput region for the uplink of a wireless network. Network Coding has emerged by allowing nodes to code over the packet traffic and was shown to increase system throughput. Linear network coding is sufficient to achieve the Max-flow Min-cut capacity [4] and for low complexity network operation, random linear network coding has been introduced in [5]. In [6] network coding was shown to have superior performance to ordinary retransmission protocols for single source multicasting to many destinations. Little work has been done to incorporate network coding with cooperation. In [7] different protocols based on deterministic network coding are proposed and shown that, for single source – single relay system with two destinations, the use of network coding at the relay increases the stable throughput. In our work we propose a network level protocol in which the relay uses the periods of silence of the source to forward its packets and hence avoiding allocating any explicit channel resources to the relay. In addition the relay performs random linear network coding on the packets it receives. We show that, compared with ARQ or protocols based on network coding [6], stable throughput for the source increases by relaying and further increase can be achieved by using network coding at the relay in the case where relay destination channel is better than source destination channels.

The paper is organized as follows. We discuss the system model in section II and introduce various protocols in section III. In section IV, we evaluate the maximum stable throughput rate of different protocols and quantify the improvements due to cooperation and network coding. In section V, we present the numerical results and in section VI we conclude the paper.

## II. SYSTEM MODEL

We consider one source node transmitting packets to each of $n$ receivers with the help of a relay as shown in Figure 1. We consider a slotted synchronous system in which each packet transmission takes one time slot. Packets are independently generated according to a Bernoulli process with average rate $\lambda$ and are addressed to each of $n$ receivers. Noise at the receivers and at the relay is assumed to be i.i.d. complex Gaussian Random process with zero mean and variance $N_o$. All links are subject to i.i.d. flat fading coefficients $h_{ij}$ which are zero mean, circularly symmetric complex Gaussian with unit variance. ACK and NACK packets are assumed to be instantaneous and error free. We adopt the SINR model for reception in which a node $j$ can successfully decode the packet transmitted by node $i$ if the SINR at node $j$ exceeds some threshold $\beta$. This can be expressed in terms of success probabilities $f_{i,j}$ over *(i-j)* link where

$$f_{ij} = \Pr\left[\frac{|h_{ij}|^2 P}{N_o} > \beta\right]$$ and $P$ is the transmission power of node $i$. If the arrival and service

processes at a queue are jointly stationary, then the queue is stable if the average arrival rate is less than the average service rate according to Loynes' theorem [8].

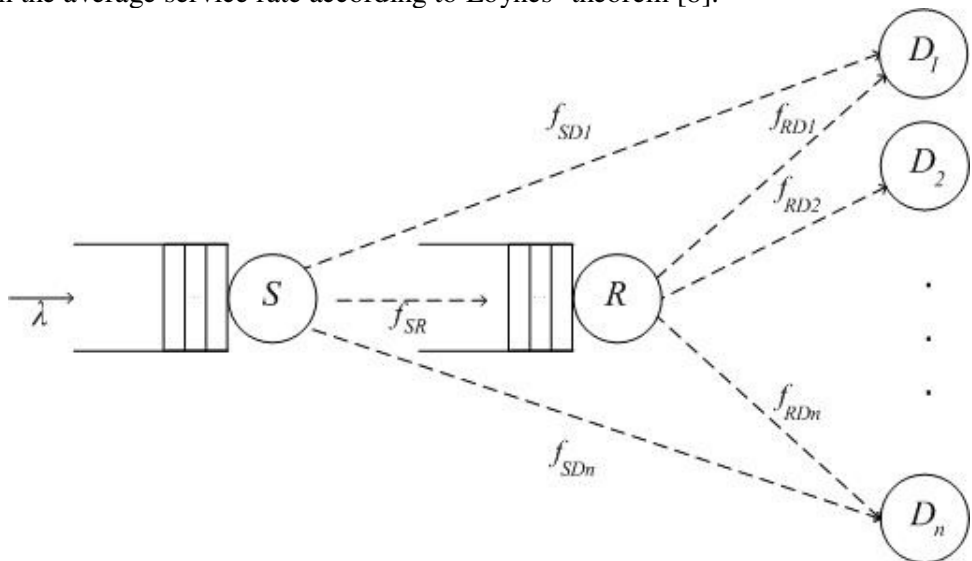

Figure 1.    System Model

## III. NETWORK PROTOCOLS

In this section we present different network protocols that can be used for multicasting networks and compare their stable throughput in section IV.

### A. Plain retransmission policy (PRP):

The transmitter node uses channel feedback and transmits a new packet only if the previous packet has been successfully received by all receivers, otherwise the same packet is retransmitted. Relay does not help the source in transmitting its traffic.

### B. Pure Random Linear Network Coding:

The transmitter node accumulates packets into groups of fixed size $K$ and transmits one random linear combination of these packets at a time until they are all successfully decoded by all the receivers, and then transmits a new combination and so on.

*C. Pure Cooperation:*

The transmitter transmits its traffic with the help of the relay but without performing any network coding on the packets. At one time slot, if the packet transmitted is successfully received by all $n$ destinations or by the relay; it is released from the source queue, otherwise it is kept in the source queue for retransmission in the following time slot. At the beginning of every time slot, the relay senses the channel. If the source does not have any traffic to send - which happens infinitely often - the relay uses these idle time slots to transmit the packets in its queue to the destinations. Hence, no explicit channel resources are accorded to the relay. A packet is released from relay queue if it is successfully decoded by all the destinations.

*D. Cooperation with Network Coding at the Relay:*

Similar to Protocol (*C*) but the relay sends linear combination of the packets it receives while the source does not perform any Network Coding.

<center>IV. STABLE THROUGHPUT ANALYSIS</center>

*A. Plain retransmission policy (PRP):*

Stability condition is given by:

$$\lambda < \left(1 + \sum_{t=1}^{\infty}\left(1 - \prod_{i=1}^{n}\sum_{r=1}^{t}(f_{SDi})(1-f_{SDi})^{r-1}\right)\right)^{-1} \tag{1}$$

The detailed proof of (1) can be found in Appendix (A).

*B. Pure Random Linear Network Coding:*

In [6] it has been shown that that the maximum stable throughput rate is given by:

$$\mu = \frac{K}{\sum_{l=K}^{\infty}\frac{1}{q^{l-K}}\left(1-\frac{1}{q^{K}}\right)\prod_{k=l+1-K}^{l-1}\left(1-\frac{1}{q^{k}}\right)\left(l + \sum_{t=l}^{\infty}\left(1 - \prod_{i=1}^{n}\sum_{r=l}^{t}\binom{r-1}{l-1}(1-f_{SDi})^{r-l}f_{SDi}^{l}\right)\right)} \tag{2}$$

Where:

$q$ : Field size used for network coding.

$K$: Number of packets over which network coding is performed.

*C. Pure Cooperation:*

System is stable if both queues (source and relay) are stable. In the following we analyze each queue separately.

*1) Source Queue:*

The arrival process at the source node is stationary by assumption. The service process is also stationary as it depends only on the source-destination channels and the source-relay channel which are subject to i.i.d. (and hence stationary) fading. Moreover, the service at the source node is independent of the arrivals and thus they are jointly stationary and we can apply Loynes' theorem at the source node. In other words, source queue is stable if $\lambda < \mu$.

Where

$$\mu = \frac{1}{1 + \sum_{t=1}^{\infty}(1-f_{SR})^{t}\left\{1 - \prod_{i=1}^{n}\sum_{r=1}^{t}(f_{SDi})(1-f_{SDi})^{r-1}\right\}} = \frac{1}{1 + \sum_{t=1}^{\infty}(1-f_{SR})^{t}\left\{1 - \prod_{i=1}^{n}\left[1-(1-f_{SDi})^{t}\right]\right\}} \tag{3}$$

Note that this is identical to Protocol (A) if $f_{SR} = 0$ as expected.

The detailed proof of (3) can be found in Appendix (B).

*2) Relay Queue:*

At the time when relay starts transmitting the packets it has in queue, there are $2^n$ possible states of the $n$ destinations regarding the state of success of packet reception. One of them is that all destinations received the packet which is uninteresting as relay is useless in that case. Thus we need to consider each of the *($2^n$ -1)* cases alone. According to our assumptions, relay would have received ACK packets from destinations that already received the packet, so the exact state of the $n$ destinations is known at the relay.

Each state of the $n$ destinations at the time when relay transmits is identified by a set (*S*) whose elements are nodes that already received the packet while source was transmitting and a set (*F*) = $S^c$ ={1,2,....n} \ S, representing nodes that failed to receive the packet and relay has to forward the packets into.

The arrival and service processes at the relay are stationary as they are functions of stationary process which is the fading process. However, they are not independent as in the case of the source. The reason behind it is that if the n-th packet gets a longtime to reach the relay which means that it has a long interarrival times, it is more likely to be successfully delivered to a larger number of destinations by the source and hence will get served faster at the relay. However, arrival and service processes are still jointly stationary because if we look at the n-th and the (n+1)st packets, then they are subject to the same conditions which depend on the i.i.d. fading and hence they are jointly stationary and hence Loynes' theorem can be applied at the relay node. Hence, the stability condition for the relay node is $\lambda_R < \mu_R$.

Where

$$\lambda_R (Upper\ bound) = \left(\frac{\lambda}{\mu}\right) f_{SR} \left(1 - \prod_{i=1}^{n} f_{SDi}\right) \tag{4}$$

$$\mu_R = \frac{1}{E[T]} = \frac{1-\rho}{E[T_R] - \rho} = \frac{\left(1 - \frac{\lambda}{\mu}\right)}{E[T_R] - \frac{\lambda}{\mu}} \tag{5}$$

$$E[T_R] = \sum_{all\ states\ S,F} \left[ \sum_{i=0}^{\infty} \left\{ 1 - \prod_{i \in F} \left[1 - (1 - f_{RDi})^i\right] \right\} \sum_{m=1}^{\infty} \left\{ f_{SR} \left(1 - \prod_{i=1}^{n} f_{SDi}\right) \left[1 - f_{SR} \left(1 - \prod_{i=1}^{n} f_{SDi}\right)\right]^{m-1} \left[ \prod_{i \in F} (1 - f_{SDi})^m \prod_{j \in S} \left[1 - (1 - f_{SDj})^m\right] \right] \right\} \right] \tag{6}$$

And $E[T_R]$ is the average number of time slots needed for the relay to deliver the packet to all destinations that failed to receive it if we allow the relay to transmit continuously, and is given by equation (6).

The detailed proof of equations (4), (5) and (6) is given in Appendix (C).

From Appendix (D) we have that the maximum stable throughput rate satisfies:

$$\lambda < \min \left( \mu, \mu \frac{\left(1 + f_{SR}\left(1 - \prod_{i=1}^{n} f_{SDi}\right) E[T_R]\right) - \sqrt{\left(1 + f_{SR}\left(1 - \prod_{i=1}^{n} f_{SDi}\right) E[T_R]\right)^2 - 4 f_{SR}\left(1 - \prod_{i=1}^{n} f_{SDi}\right)}}{2 f_{SR}\left(1 - \prod_{i=1}^{n} f_{SDi}\right)} \right) \tag{7}$$

Where $\mu$ is given by (3).

*D. Cooperation with NC at the relay:*

The arrival and service rates of the source node are as before as well as the arrival rate to the relay. The only difference is the service rate of the relay.

The stability condition for Protocol (D) is:

$$\lambda < \min\left(\mu, \mu \frac{\left(K + f_{SR}\left(1 - \prod_{i=1}^{n} f_{SDi}\right)E[T_R]\right) - \sqrt{\left(K + f_{SR}\left(1 - \prod_{i=1}^{n} f_{SDi}\right)E[T_R]\right)^2 - 4Kf_{SR}\left(1 - \prod_{i=1}^{n} f_{SDi}\right)}}{2f_{SR}\left(1 - \prod_{i=1}^{n} f_{SDi}\right)}\right)$$

$$\tag{8}$$

$$E[T_R] = \sum_{all\ states\ F_1,...F_K} E[T_R \mid State\ F_1, F_2, ...F_K]\Pr[State\ F_1, F_2, ...F_K] \tag{9}$$

$$E[T_R \mid State\ F_1, F_2, ...F_K] = \sum_{l=K}^{\infty} \frac{1}{q^{l-K}}\left(1 - \frac{1}{q^K}\right)\prod_{k=l+1-K}^{l-1}\left(1 - \frac{1}{q^k}\right)\left(l + \sum_{t=l}^{\infty}\left(1 - \prod_{\substack{i \in \bigcup_{j=1}^{K} F_j}}\sum_{r=l}^{t}\binom{r-1}{l-1}(1 - f_{RDi})^{r-l}f_{RDi}^{\ l}\right)\right) \tag{10}$$

$$\Pr[State\ F_1, F_2, ...F_K] = \prod_{l=1}^{K}\left[\sum_{m=1}^{\infty}\left\{f_{SR}\left(1 - \prod_{i=1}^{n} f_{SDi}\right)\left[1 - f_{SR}\left(1 - \prod_{i=1}^{n} f_{SDi}\right)\right]^{m-1}\left[\prod_{i \in F_l}(1 - f_{SDi})^m \prod_{j \in S_l}\left[1 - (1 - f_{SDj})^m\right]\right]\right\}\right] \tag{11}$$

The stability condition for Protocol (*D*) is given by equations (8), (9), (10), (11).
Where:
$\mu$ is as given by equation (3).
$q$ : Field size of Network Coding used at the relay.
$K$ : Number of packets over which encoding is performed at the relay.
$F_i$, $S_i$ describe the state of reception of the $i$-th packet at the $n$ destinations at the time when relay begins transmitting.
The detailed proof of equations (8), (9), (10), (11) is given in Appendix (E).

## V. NUMERICAL RESULTS

In Figure 2, we compare the maximum stable throughput rates of the protocols (*A*), (*B*) and (*C*). For clarity of presentation, we consider a symmetric configuration in which all the source-destinations links have the same success probabilities which we will denote by $p$ for all the protocols. For protocol (*B*), $K$ and $q$ are respectively the field size and number of packets over which we perform Network coding. For protocol (*C*), we denote by $pr$ the success probability of the relay-destinations channel, and by $fsr$ the success probability of the source relay channel. As expected, cooperation increases the stable throughput rate as it overcomes deep fading over direct links resulting in faster emptying of users' queues.

In Figures 3 and 4, we compare between protocols $C$ and $D$. $K$ and $q$ are the parameters of NC used at the relay. Other parameters are same as defined before. It is clear that Random Linear NC at the relay increases the stable throughput with increasing $q$ or $K$ or both and it becomes more advantageous as the number of destinations gets larger and larger as we expect.

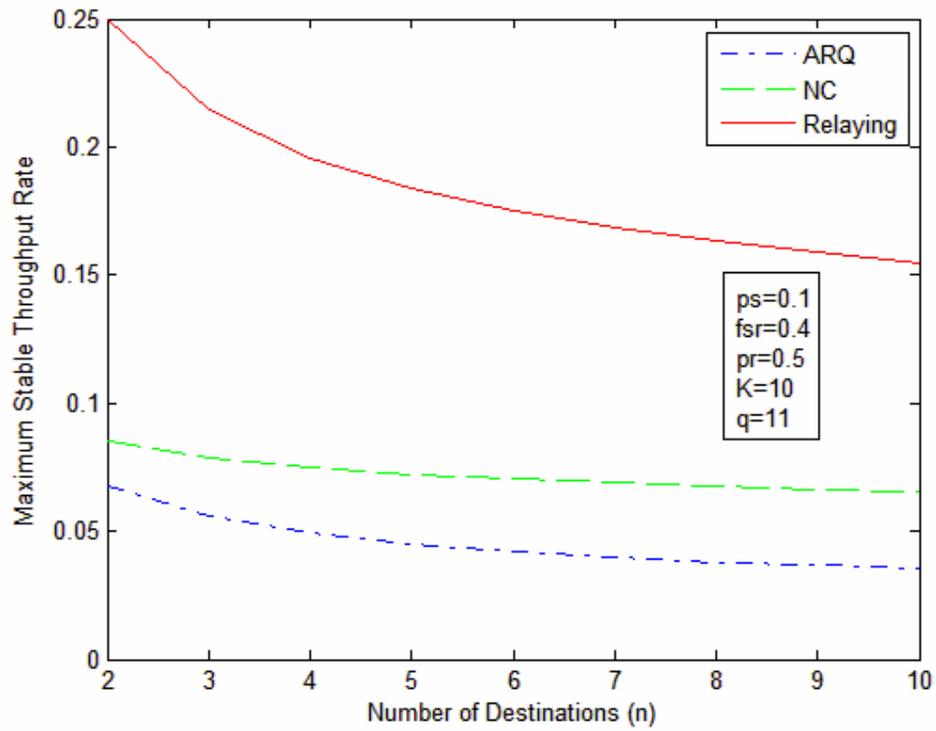

Figure 2 : Maximum Stable Throughput Rate of Protocols (A), (B), (C)

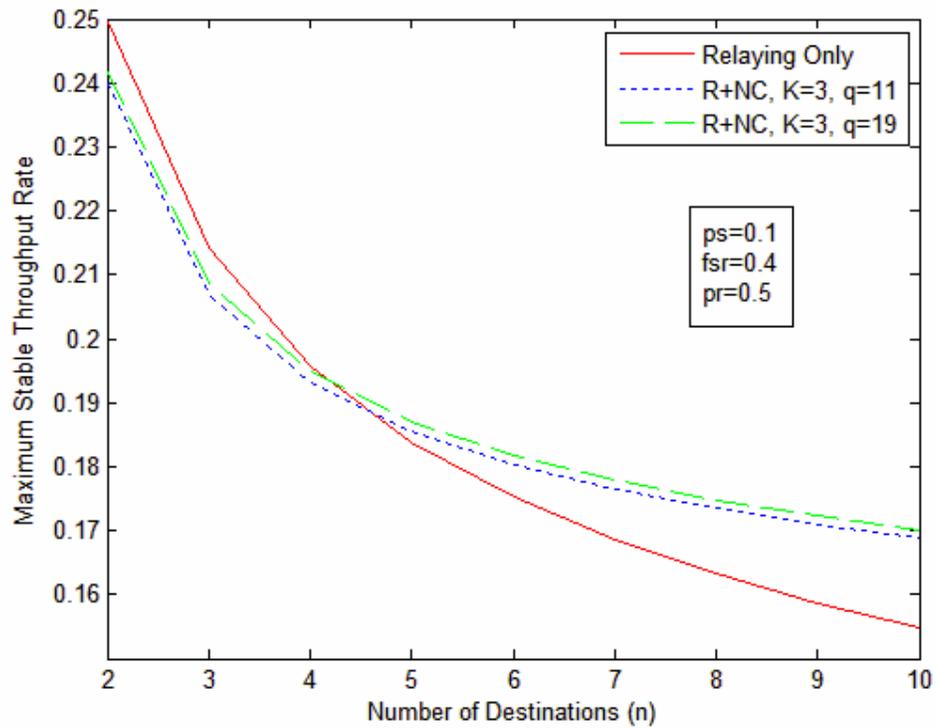

Figure 3 : Effect of using Network Coding at the Relay for various values of field size (q)

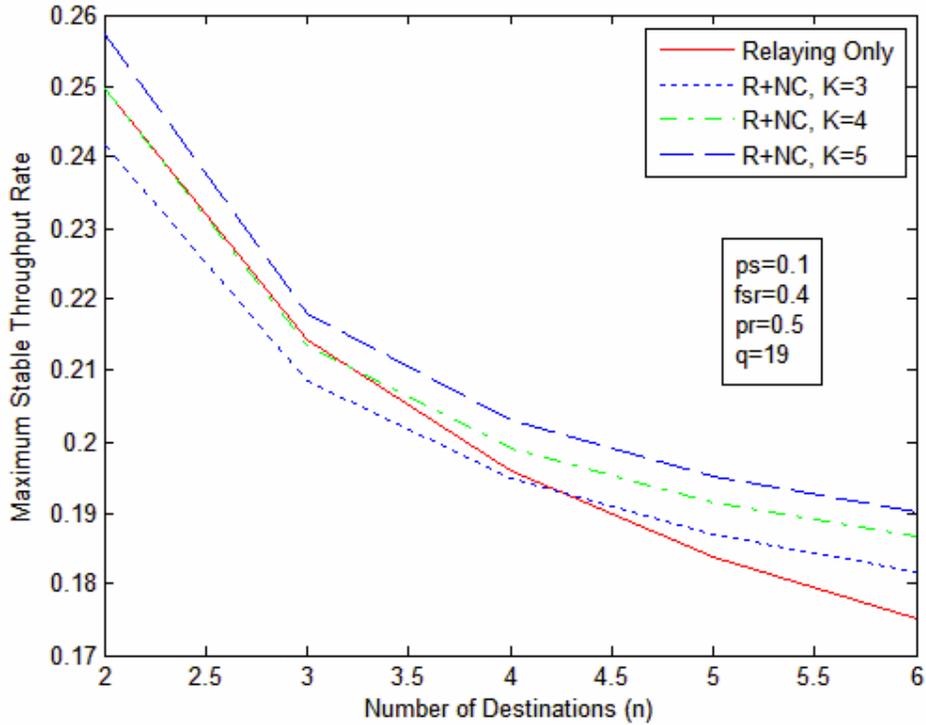

Figure 4 : Effect of using Network Coding at the Relay for various values of number of encoded packets (K)

## VI. Conclusion

In this paper, we proposed and analyzed a protocol for wireless multicasting networks that combines both cooperation at the network level and Network Coding at the relay node. The proposed protocol allows the relay to use the idle time slots of the source and hence avoids according any explicit resources to the relay. Our analysis showed that cooperation largely increased the maximum stable throughput rate and additionally, it can be further increased by using Network Coding at the relay. Future work includes analyzing the delay and energy efficiency of this protocol and generalizing to the case where relay is not a dedicated relay but a source node having its own traffic.

## Appendix A
### Proof of Stability Condition For Protocol (A)

Let $N_i$ be the number of times the source has to transmit the packet until the i-th receiver gets it, thus:

$$\Pr[N_i = k] = (f_{SDi})(1 - f_{SDi})^{k-1}, \qquad i=1,2,3,\dots,n \ ; \qquad k=1,2,3,\dots. \qquad (12)$$

Let $T$ denotes the number of times the source has to transmit until all the nodes get the packet, then:

$$T = \max_{i=1,2,\dots,n} N_i \qquad (13)$$

$$\Pr[T \le t] = \Pr[T_1 \le t, T_2 \le t,\dots,T_n \le t] = \prod_{i=1}^{n} \Pr[T_i \le t] = \prod_{i=1}^{n} \sum_{r=1}^{t} (f_{SDi})(1 - f_{SDi})^{r-1} \qquad (14)$$

$$\mathrm{E}[T] = \sum_{t=0}^{\infty} \Pr[T > t] = 1 + \sum_{t=1}^{\infty} \left(1 - \prod_{i=1}^{n} \sum_{r=1}^{t} (f_{SDi})(1 - f_{SDi})^{r-1}\right) \qquad (15)$$

Finally,
$$\mu = \frac{1}{E[N]} = \left(1 + \sum_{t=1}^{\infty}\left(1 - \prod_{i=1}^{n}\sum_{r=1}^{t}(f_{SDi})(1-f_{SDi})^{r-1}\right)\right)^{-1} \qquad (16)$$

Stability condition:
$$\lambda < \left(1 + \sum_{t=1}^{\infty}\left(1 - \prod_{i=1}^{n}\sum_{r=1}^{t}(f_{SDi})(1-f_{SDi})^{r-1}\right)\right)^{-1} \qquad (17)$$

<div align="center">

APPENDIX B

PROOF OF STABILITY CONDITION FOR SOURCE QUEUE IN PROTOCOL (C)

</div>

The source queue is served if the packet is successfully delivered at the relay or all $n$ destinations. Let $T$ be the number of time slots needed to serve the source queue, then:
$$T = \min\{T_L, T_D\} \qquad (18)$$
Where

$T_L$: Number of time slots needed to deliver the packet to the relay.

$T_D$: Number of time slots needed to deliver the packet to all '$n$' destinations.
$$\Pr[T_L = k] = f_{SR}(1-f_{SR})^{k-1}, \qquad k = 1,2,3,4,\ldots. \qquad (19)$$
$$T_D = \max_{i=1,2,\ldots n} T_{D,i} \qquad (20)$$

Where $T_{Di}$ : is the number of time slots needed to deliver the packet to the i-th destination.
$$\Pr[T_{D,i} = r] = (f_{SDi})(1-f_{SDi})^{r-1}, \qquad r = 1,2,3,\ldots \qquad (21)$$

$$\Pr[T_D \le t] = \prod_{i=1}^{n}\sum_{r=1}^{t}(f_{SDi})(1-f_{SDi})^{r-1} \qquad (22)$$

$$\Pr[T > t] = \Pr[T_D > t]\Pr[T_L > t] = \left\{1 - \prod_{i=1}^{n}\sum_{r=1}^{t}(f_{SDi})(1-f_{SDi})^{r-1}\right\}\left\{\sum_{k=t+1}^{\infty} f_{SR}(1-f_{SR})^{k-1}\right\}$$

$$= (1-f_{SR})^t\left\{1 - \prod_{i=1}^{n}\sum_{r=1}^{t}(f_{SDi})(1-f_{SDi})^{r-1}\right\} \qquad (23)$$

And
$$E[T] = 1 + \sum_{t=1}^{\infty}(1-f_{SR})^t\left\{1 - \prod_{i=1}^{n}\sum_{r=1}^{t}(f_{SDi})(1-f_{SDi})^{r-1}\right\} \qquad (24)$$

Finally,
$$\mu = \frac{1}{E[T]} = \frac{1}{1 + \sum_{t=1}^{\infty}(1-f_{SR})^t\left\{1 - \prod_{i=1}^{n}\sum_{r=1}^{t}(f_{SDi})(1-f_{SDi})^{r-1}\right\}} = \frac{1}{1 + \sum_{t=1}^{\infty}(1-f_{SR})^t\left\{1 - \prod_{i=1}^{n}\left[1-(1-f_{SDi})^t\right]\right\}}$$
$$(25)$$

<div align="center">

APPENDIX C

PROOF OF STABILITY CONDITION FOR RELAY QUEUE IN PROTOCOL (C)

</div>

An upper bound on the arrival rate for the relay is to look at only one slot, regardless of the history of transmissions at the source. Relay will have an arrival at that slot if the relay source channel is not in outage while at least one of the source destination channels is in outage.

The deviation of this approximation from the real situation can be seen by the following example:

Assume for all (N-1) slots, relay could not decode the packet while a subset (S) of the destinations could. If at the N-th slot, relay successfully decoded the packet but also all the nodes in (S$^c$) could, then this should not be counted as arrival if we look at the history of transmissions while that would be counted as an arrival by the approximation. As we are counting for more arrivals than true, we have an upper bound on the arrival rate. However, this can be taken into account, by assuming that such packet at the relay (that has been successfully received by all the destinations) will take zero time slot from the relay to get served and hence, it does not affect the actual unsuccessful packets. So, by assuming that a packet whose state at the destinations is S={1,2,....,n} takes zero time to get served, we can evaluate the actual performance of the system.

$\lambda_R(Upper\ bound) = \Pr[\{Q_S \neq 0\}]\Pr[\text{SR link is not in outage}]\Pr[\text{at least one SD links is in outage}]$.

$$\lambda_R(Upper\ bound) = \left(\frac{\lambda}{\mu}\right)f_{SR}\left(1 - \prod_{i=1}^{n} f_{SDi}\right) \tag{26}$$

Also, if $T*$ is the number of time slots until the relay has an arrival, then:

$$\Pr[T* = m] = f_{SR}\left(1 - \prod_{i=1}^{n} f_{SDi}\right)\left[1 - f_{SR}\left(1 - \prod_{i=1}^{n} f_{SDi}\right)\right]^{m-1} \tag{27}$$

Given that destinations are at a certain state before relay begins transmitting:
Let $T_R$ be the number of time slots needed for relay to serve the packet it has in its queue assuming that it continuously transmits, i.e. number of time slots needed for packet to reach all $n$ destinations. Given a certain state of the destinations, $T_R$ is the number of time slots needed for the relay to deliver the packet into all destinations in the set $F$ previously defined.

$$T_R = \max_{i \in F} T_{R,i} \tag{28}$$

Where $T_{R,i}$ is the time for the relay to reach $i$-th node, which is geometrically distributed with parameter ($f_{RDi}$).

$$\Pr[T_{R,i} = k] = f_{RDi}(1 - f_{RDi})^{k-1} \tag{29}$$

$$\Pr[T_R \leq t] = \prod_{i \in F}\sum_{k=1}^{t} f_{RDi}(1 - f_{RDi})^{k-1} = \prod_{i \in F}\left[1 - (1 - f_{RDi})^t\right] \tag{30}$$

$$E[T_R \mid certain\ state\ S, F] = \sum_{t=0}^{\infty}\left\{1 - \prod_{i \in F}\left[1 - (1 - f_{RDi})^t\right]\right\} \tag{31}$$

Thus

$$E[T_R] = \sum_{all\ states\ S, F}\sum_{t=0}^{\infty}\left\{1 - \prod_{i \in F}\left[1 - (1 - f_{RDi})^t\right]\right\}\Pr[Destinations\ are\ in\ state\ S, F\ when\ relay\ begins\ trnasmiting] \tag{32}$$

Now:

$\Pr[Destinations\ are\ in\ state\ S, F\ when\ relay\ begins\ trnasmiting] =$

$= \sum_{m=1}^{\infty}\Pr[\text{Destinations are in state S,F when relay starts transmission| relay has arrival after exactly 'm' transmissions]*Pr[Relay has arrival after exactly 'm' transmissions}]$

$= \sum_{m=1}^{\infty}\Pr[\text{All nodes in } S \text{ decoded the packet within 'm' trials while all nodes in } F \text{ couldn't decode the packet in 'm' trials}]\Pr[T_R = m] \tag{33}$

And,
$\Pr[\text{All nodes in } S \text{ decoded the packet within 'm' trials while all nodes in } F \text{ couldn't decode the packet in 'm' trials}] = \prod_{i \in F}(1 - f_{SDi})^m \prod_{j \in S}\left[1 - (1 - f_{SDj})^m\right] \tag{34}$

Hence by using (32),

$$E[T_R] = \sum_{all\ states\ S,F} \left[ \sum_{r=0}^{\infty} \left\{ 1 - \prod_{i \in F} \left[ 1 - (1 - f_{RDi})^r \right] \right\} \sum_{m=1}^{\infty} \left\{ f_{SR} \left( 1 - \prod_{i=1}^{n} f_{SDi} \right) \left[ 1 - f_{SR} \left( 1 - \prod_{i=1}^{n} f_{SDi} \right) \right]^{m-1} \left[ \prod_{i \in F} (1 - f_{SDi})^m \prod_{j \in S} \left[ 1 - (1 - f_{SDj})^m \right] \right] \right\} \right]$$

(35)

Let $v_1, v_2, \ldots$ be a sequence of random variables representing the number of successive time slots in which the source is to be busy, then it's clear that this sequence represent an i.i.d sequence.

For certain time slot:

$\Pr[an\ arrival\ occurs] = 1 - \Pr[no\ arrivals] = \lambda$  (36)

The source queue is a Geo/G/1 queue. Thus using the result in [9]:

$$E[Busy\ period] = \frac{1}{\mu - \lambda}$$  (37)

$$E[v] = \lambda E[Busy\ period] = \left( \frac{\lambda}{\mu} \right) \Big/ \left( 1 - \frac{\lambda}{\mu} \right)$$  (38)

Let $T$ be the total number of time slots needed for the relay to get served including those in which the source will be transmitting, then we have:

$$T = T_R + \sum_{i=1}^{T_R - 1} v_i$$  (39)

Because between $T_R$ relay time slots we can have $T_R$ -1 source busy periods (possibly of length zero).

$$E[T] = E[T_R] + (E[T_R] - 1)E[v] = E[T_R] + (E[T_R] - 1) \frac{\frac{\lambda}{\mu}}{1 - \frac{\lambda}{\mu}} = \frac{E[T_R]}{1 - \frac{\lambda}{\mu}} - \frac{\frac{\lambda}{\mu}}{1 - \frac{\lambda}{\mu}}$$

Let $\rho = \frac{\lambda}{\mu}$

$$E[T] = \frac{E[T_R] - \rho}{1 - \rho}$$  (40)

Thus:

$$\mu_R = \frac{1}{E[T]} = \frac{1 - \rho}{E[T_R] - \rho} = \frac{\left( 1 - \frac{\lambda}{\mu} \right)}{E[T_R] - \frac{\lambda}{\mu}}$$  (41)

APPENDIX D

PROOF OF MAXIMUM STABLE THROUGHPUT RATE IN PROTOCOL (C)

The system is stable if the source and relay queues are stable.

By Appendices (B) and (C), for stability we should have:

(1) $\lambda < \mu$  (42)

Or equivalently: $\rho < 1$

And:

(2) $\rho f_{SR} \left( 1 - \prod_{i=1}^{n} f_{SDi} \right) < \frac{(1 - \rho)}{E[T_R] - \rho}$

Or equivalently:

$$\rho^2 f_{SR}\left(1-\prod_{i=1}^{n} f_{SDi}\right) - \rho f_{SR}\left(1-\prod_{i=1}^{n} f_{SDi}\right)E[T_R] > \rho - 1$$

So:

$$\rho^2 f_{SR}\left(1-\prod_{i=1}^{n} f_{SDi}\right) - \rho\left(1 + f_{SR}\left(1-\prod_{i=1}^{n} f_{SDi}\right)E[T_R]\right) + 1 > 0$$

- Now looking at the equation:

$$\rho^2 f_{SR}\left(1-\prod_{i=1}^{n} f_{SDi}\right) - \rho\left(1 + f_{SR}\left(1-\prod_{i=1}^{n} f_{SDi}\right)E[T_R]\right) + 1 = 0 \qquad (43)$$

This is a quadratic equation having 2 positive roots. The product of the roots is $\dfrac{1}{f_{SR}\left(1-\prod_{i=1}^{n} f_{SDi}\right)}$

Which is always larger than one, thus we have at least one root larger than one.
Thus $\rho$ must be smaller than the smallest root or larger than the largest root, but also $\rho < 1$,
Thus:
$\rho < \min\{1, \text{smallest root of equation (43)}\}$
Smallest root =

$$\frac{\left(1 + f_{SR}\left(1-\prod_{i=1}^{n} f_{SDi}\right)E[T_R]\right) - \sqrt{\left(1 + f_{SR}\left(1-\prod_{i=1}^{n} f_{SDi}\right)E[T_R]\right)^2 - 4 f_{SR}\left(1-\prod_{i=1}^{n} f_{SDi}\right)}}{2 f_{SR}\left(1-\prod_{i=1}^{n} f_{SDi}\right)} \qquad (44)$$

So:

$$\rho < \min\left\{1, \frac{\left(1 + f_{SR}\left(1-\prod_{i=1}^{n} f_{SDi}\right)E[T_R]\right) - \sqrt{\left(1 + f_{SR}\left(1-\prod_{i=1}^{n} f_{SDi}\right)E[T_R]\right)^2 - 4 f_{SR}\left(1-\prod_{i=1}^{n} f_{SDi}\right)}}{2 f_{SR}\left(1-\prod_{i=1}^{n} f_{SDi}\right)}\right\} \qquad (45)$$

APPENDIX E
PROOF OF STABILITY CONDITION FOR PROTOCOL (D)

For every packet of the $K$ packets, we have a corresponding state at the $n$ destinations, i.e. for every packet $j=1, 2..., K$ we have a set $(F_j)$ representing the destinations that didn't receive that packet. For the relay to serve the $K$ packets, all the destinations in the union of the sets $F_j$ (i.e. $\bigcup_{j=1}^{K} F_j$) must be able to successfully decode the $K$ packets. Thus, given a certain state $(F_1, F_2 ...F_K)$ of the $K$ packets, using this in the result of Protocol (B) in [6] we have:

$$E[T_R \mid State\ F_1, F_2, ...F_K] = \sum_{l=K}^{\infty} \frac{1}{q^{l-K}}\left(1 - \frac{1}{q^K}\right)\prod_{k=l+1-K}^{l-1}\left(1 - \frac{1}{q^k}\right)\left(l + \sum_{t=l}^{\infty}\left(1 - \prod_{i\in \bigcup_{j=1}^{K} F_j} \sum_{r=l}^{t}\binom{r-1}{l-1}(1-f_{RDi})^{r-l} f_{RDi}^{l}\right)\right) \qquad (46)$$

Also, following similar analysis as in Appendix (C), and noting the independence between the events of having different packets in different states, we get:

$$\Pr\left[State\ F_1, F_2,...F_K\right] = \prod_{l=1}^{K}\left[\sum_{m=1}^{\infty}\left\{f_{SR}\left(1-\prod_{i=1}^{n}f_{SDi}\right)\left[1-f_{SR}\left(1-\prod_{i=1}^{n}f_{SDi}\right)\right]^{m-1}\left[\prod_{i\in F_l}\left(1-f_{SDi}\right)^m\prod_{j\in S_l}\left[1-\left(1-f_{SDj}\right)^m\right]\right]\right\}\right]$$

(47)

Finally:

$$E[T_R] = \sum_{all\ states\ F_1,...F_K} E[T_R\mid State\ F_1, F_2,...F_K]\Pr\left[State\ F_1, F_2,...F_K\right]$$

(48)

Note that $E[T_R]$ represents the number of time slots needed to serve the $K$ packets if the relay continuously transmits (i.e. not using only the idle time periods of the source).

From Appendix (C) we have that:

$$E[T] = \frac{E[T_R]-\rho}{1-\rho}$$

Thus the number of time slots needed to serve one packet is on average equal to:

$$E[T*] = \frac{E[T_R]-\rho}{K-K\rho}$$

Thus stability is obtained if:

$$\rho\, f_{SR}\left(1-\prod_{i=1}^{n}f_{SDi}\right) < \frac{K(1-\rho)}{E[T_R]-\rho}$$

-By similar analysis as in Appendix (D):

$$\lambda < \min\left(\mu, \mu\frac{\left(K+f_{SR}\left(1-\prod_{i=1}^{n}f_{SDi}\right)E[T_R]\right)-\sqrt{\left(K+f_{SR}\left(1-\prod_{i=1}^{n}f_{SDi}\right)E[T_R]\right)^2-4Kf_{SR}\left(1-\prod_{i=1}^{n}f_{SDi}\right)}}{2f_{SR}\left(1-\prod_{i=1}^{n}f_{SDi}\right)}\right)$$

(49)

REFERENCES

[1] A. Sendonaris, E. Erkip, and B. Aazhang, "User cooperation diversity-Part I: System description," *IEEE Trans. Commun.*, vol. 51, pp. 1927–1938, Nov. 2003.

[2] J. N. Laneman, D. N. C. Tse, and G.W.Wornell, "Cooperative diversity in wireless networks: efficient protocols and outage behavior," *IEEE Trans. Inform. Theory*, vol. 50, no. 12, pp. 3062–3080, Dec. 2004.

[3] A. K. Sadek, K. J. R. Liu and A. Ephremides, "Collaborative Multiple Access Protocols for Wireless Networks: Protocol Design and Stability Analysis," *Proceedings IEEE Conference on Information Sciences and Systems (CISS)*, Princeton, NJ, 2006.

[4] S. Y. R. Li, R. W. Yeung, and N. Cai, "Linear Network Coding," *IEEE Trans. Inform. Theory*, vol. 49, no. 2, pp. 1204 1216, Feb. 2003.

[5] T. Ho, R. Koetter, M. Medard, D. R. Karger, and M. Effros, "The Benefits of Coding over Routing in a Randomized Setting," in *Proc. IEEE Int. Symp. Information Theory*, pp. 440, Yokohama, Japan, June 2003.

[6] Y. E Sagduyu and A.Ephremides, "On Network Coding for Stable Multicast Communication," *IEEE Military Communications Conference MILCOM*, Oct 2007.

[7] P. Fan, C. Zhi, C. Wei and K. Ben Letaief, "Reliable relay assisted wireless multicast using network coding," *IEEE J. Select. Areas Commun.*, vol . 27, pp. 749-762, Jun. 2009.

[8] R. Loynes, "The Stability of a Queue with Non-interdependent Inter-arrival and Service Times," *Proc. Camb. Philos. Soc.*, vol. 58, pp. 497-520, 1962.

[9] H. Takagi, *"Queueing Analysis - Vol3"*, North-Holland, 1993.